\documentclass{emulateapj}
\usepackage{amssymb}   
\usepackage{colordvi}   

\shorttitle{Stellar populations across the BH mass -- velocity-dispersion relation} 
\shortauthors{I. Mart\'in-Navarro et al. }

\def\gsim{ \lower .75ex \hbox{$\sim$} \llap{\raise .27ex \hbox{$>$}} }
\newcommand{\msun}{\hbox{M$_{\odot}$}}

\begin{document}
\title{Stellar populations across the black hole mass -- velocity dispersion relation}

\author{Ignacio Mart\'in-Navarro$^{1}$, Jean P. Brodie$^1$, Remco C. E. van den Bosch$^2$, \break 
Aaron J. Romanowsky$^{3,1}$ \& Duncan A. Forbes$^4$ }
\affil{$^{1}$University of California Observatories, 1156 High Street, Santa Cruz, CA 95064, USA}
\affil{$^{2}$Max-Planck Institut f\"ur Astronomie, Konigstuhl 17, D-69117 Heidelberg, Germany}
\affil{$^{3}$Department of Physics and Astronomy, San Jos\'e State University, One Washington Square, San Jose, CA 95192, USA}
\affil{$^{4}$Centre for Astrophysics and Supercomputing, Swinburne University, Hawthorn VIC 3122, Australia}
\email{email: imartinn@ucsc.edu}

\begin{abstract}
Coevolution between supermassive black holes (BH) and their host galaxies is universally adopted in models for galaxy formation. In the absence of feedback from active galactic nuclei, simulated massive galaxies keep forming stars in the local Universe. From an observational point of view, however, such coevolution remains unclear. We present a stellar population analysis of galaxies with direct BH mass measurements and the BH mass--$\sigma$  relation as a working framework. We find that over-massive BH galaxies, i.e., galaxies lying above the best-fitting BH mass--$\sigma$ line, tend to be older and more $\alpha$-element enhanced than under-massive BH galaxies. The scatter in the BH mass--$\sigma$--[$\alpha/$Fe] plane is significantly lower than in the standard  BH mass--$\sigma$ relation. We interpret this trend as an imprint of active galactic nucleus feedback on the star formation histories of massive galaxies.
\end{abstract}


\section{Introduction}

The suppression of star formation via active galactic nucleus (AGN) feedback plays a crucial role in state-of-the-art numerical simulations \citep{Vogelsberger,Schaye}, but its observational effects are difficult to establish. The distribution of AGNs in the color--magnitude plane \citep{Martin07,Schawinski07} has been traditionally used as an indirect method to empirically constrain any AGN effect in nearby galaxies. With a detailed chemical evolution treatment in the most recent cosmological simulations \citep[e.g.][]{Crain15}, a new window of opportunity opens for understanding the effect of AGN on nearby objects. Based on the Evolution and Assembly of GaLaxies and their Environments (EAGLE) cosmological simulations, \citet{Segers} have recently linked AGN feedback to the over-abundance of $\alpha$-elements in massive galaxies -- a well-known property of nearby early-type galaxies \citep{Thomas,dlr11,Conroy14}.

The existence of an [$\alpha$/Fe]--galaxy mass relation suggests a link between the star formation time-scale of a galaxy and its mass. Whereas $\alpha$-elements are produced in core collapse supernovae (SNe) with very short life-times, the onset of Type Ia SNe occurs later ($\sim 1$~Gyr), releasing mainly iron to the interstellar medium. Therefore the relative abundance of $\alpha$-elements to iron reflects how long SNe Ia have been able to pollute the medium: the less  $\alpha$-enhanced a stellar population is, the more extended has been its star formation. In the scenario proposed by \citet{Segers}, more massive galaxies host and fuel more massive black holes (BHs) in their centers, leading to a stronger AGN effect which ultimately quenches the star formation more rapidly. The [$\alpha$/Fe]--galaxy mass relation would appear therefore as a natural consequence of the coevolution between BHs and galaxies. 

However, the (level of) coevolution is still under debate. The BH masses do correlate with the host galaxies properties \citep[see][for a review]{kh13}. The tightest correlation is the so-called M--$\sigma$ relation which links the BH mass and stellar velocity dispersion ($\sigma$) of the host galaxy \citep{Beifiori,vdB}. Whether this BH mass--$\sigma$ relation results from a causal connection \citep[e.g.][]{Silk98,Fabian,King} or not \citep{Peng07,Jahnke11} remains an open question.

The effect of the AGN feedback depends strongly on the BH accretion rate. Close to the Eddington limit, the amount of energy radiated around the BH is expected to be large enough to effectively quench the ongoing star formation \citep{Fabian12}. This so-called quasar mode would take place at high redshifts ($z \sim 2-3$), and precedes the less energetic maintenance mode, which happens at lower BH accretion rates. The maintenance mode is thought to be responsible for continuously heating the gaseous halos around nearby massive galaxies. The properties of present-day massive galaxies would then be a combination of the early quenching associated with the quasar mode, and the more extended maintenance mode which inhibits further star formation \citep{Voit,Choi}.
 
However, observational evidence of AGN feedback is inconclusive. On the one hand, strong nuclear outflows have been reported in a wide variety of environments, from luminous quasars \citep{Greene11} to nearby quiescent galaxies \citep{Cheung}. Moreover, \citet{Beifiori} have tentatively explored the connection between nuclear activity and the M--$\sigma$ relation. On the other hand, and contrary to what would be expected from negative AGN feedback, the most (X-ray) luminous AGNs are found in strongly star forming galaxies \citep{Rovilos}. 

In this Letter we report our attempts to quantify the effect of the central BH on the evolution of the host galaxy. We conduct a stellar population analysis of galaxies with known BH masses, finding significant differences in the stellar population properties depending on the location of the galaxy in the BH mass--$\sigma$ plane. This suggests a strong degree of coevolution between galaxies and their central BHs.

\section{Black hole data}

We use as a reference the BH sample of \citet{vdB}, which provides not only a large (230 objects) and updated compilation of BH masses and $\sigma$, but also consistent measurements of the light profile concentration $C_{28} \equiv 5 \log R_{20} / R_{80}$. The stellar population properties were extracted for galaxies
belonging to the Hobby-Eberly Telescope Massive Galaxy Survey \citep[HETMGS,][]{hetmgs}. HETMGS consists of a long-slit spectroscopic survey of 1022 objects, using the Marcario Low Resolution Spectrograph on the 10~m HET, at intermediate spectral resolution (4.8 and 7.5 \AA) depending on the slit width (1'' and 2'', respectively). Each object in the sample was observed for at least 15 min, covering a spectral range from 4300 to 7400 \AA.

After visual inspection, we removed from the sample those objects with poor kinematical fits (Signal-to-Noise $<5\,-\,10$), and also those galaxies where the strength of the emission lines dominated the central spectrum (Amplitude-over-Noise $>$ 50), thereby avoiding strong contamination of the Balmer absorption lines, our main age indicator. Finally, we also excluded from the analysis galaxies with recession velocities $z>0.033$, since prominent telluric lines could affect our stellar population analysis. Although galaxies with velocity dispersions down to $\log\sigma=1.8$ (km~s$^{-1}$) fulfilled all these criteria, the vast majority of objects ranged from $\log\sigma=2$ to $\log\sigma=2.5$. Moreover, for galaxies below the L$^*$ characteristic luminosity, the star formation efficiency is mainly regulated by stellar feedback \citep{Silk}. Thus, to isolate the effect of the AGN in our sample, we limited our study to galaxies with $\log\sigma>2$.

\begin{figure}
 \begin{center}
  \includegraphics[scale=0.50]{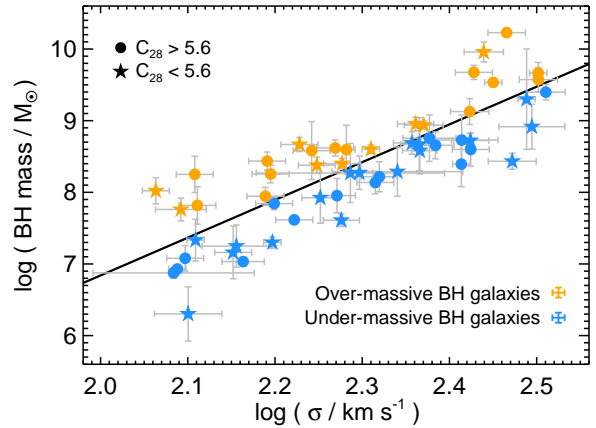}
  \caption{The BH mass -- velocity dispersion relation for our sample of galaxies. Orange and blue dots correspond to over- and under-massive BH galaxies, i.e., galaxies above and below the best-fitting relation of \citet{vdB} (black solid line). Objects with highly concentrated brightness profiles ($C_{28} > 5.6$) are shown as filled circles, whereas filled stars correspond to lower ($C_{28} > 5.6$) concentration index.}
  \label{sigma}
 \end{center}
\end{figure}

Our working sample consists then of 57 galaxies, with BH mass, $\sigma$, and $R_{\rm e}$ measurements, plus long-slit spectroscopic data to perform the stellar population analysis. Fig.~\ref{sigma} shows the BH mass -- $\sigma$ relation for the sample, where different symbols indicate different concentration indices. The threshold value $C_{28} = 5.6$ corresponds to the median of the distribution.

\section{Stellar population analysis}

The stellar population analysis is based on the latest version of the MILES models \citep{miles,alpha}, which range from 0.03 to 14 Gyr in age, from $-2.27$ to $+0.40$ dex in metallicity, and from $+0.0$ to $+0.4$ dex in [$\alpha$/Fe]. The slope of the stellar initial mass function was assumed to have the Milky Way value \citep[$\Gamma_\mathrm{b} = 1.3$ in the MILES notation;][]{kroupa}. All our measurements are luminosity-weighted, single stellar population (SSP) equivalent values.

To perform a homogeneous stellar population analysis we extracted the central spectrum of each galaxy within a fixed aperture of $R_{\rm e}/8$.  We then used the Penalized Pixel-Fitting code \citep[pPXF;][]{ppxf} to simultaneously fit the kinematics and remove the nebular emission.

Although the information in line-strength indices is mainly encoded in the depth of the spectral features, index measurements are also sensitive to the shape of the continuum. Therefore, before attempting a detailed line-strength analysis, the HETMGS data must be flux calibrated. To do so, we first obtain a zero-order estimate of the stellar population parameters by fitting the prominent H$\,\beta$, Mgb5170, Fe5270 and Fe5335 lines after removing their continuum. This leads to approximate age, metallicity, and [Mg/Fe] values, which are used to estimate a best-fitting template. We then divide the observed spectrum by the best-fitting template, fitting this ratio by a low (4$^\mathrm{th}$ grade) order polynomial which is finally used to correct the continuum. All our analysis is based on these continuum-corrected spectra.

To derive the abundance ratios we follow the approach described in \citet[][\S8.2.4]{alpha}, which consists of two basic steps. First, the mean (luminosity-weighted) age is measured using a standard H$_{\beta o}$--[MgFe]$'$ diagram\footnote{We used the H$_{\beta o}$ and [MgFe]$'$ indices as defined in \citet{hbo} and \cite{mgfe}, respectively.}. Second, the [Mg/Fe] value is calculated as the difference between the Mg and Fe metallicities, corrected by a constant factor. More explicitly, our [Mg/Fe] ratio is given by 

\begin{equation}
 \mathrm{[Mg/Fe]=C\times(\,[M/H]_{Mgb}-[M/H]_{\langle Fe\rangle}})
\end{equation}

\noindent
where $\mathrm{[M/H]_{Mgb}}$ and $\mathrm{[M/H]_{\langle Fe\rangle}}$ are the best-fitting metallicities measured at fixed age, measured from the Mgb and $\mathrm{\langle Fe\rangle}$ line-strength indices, respectively. The constant factor, $\mathrm{C}$, depends on the set of models; for the MILES solar scale models, based on the BaSTI set of isochrones \citep{Basti1,Basti2}, it has a value of 0.59 \citep{alpha}. 

In summary, we used the H$\,\beta$ and [MgFe]$'$ line-strength indices to constrain ages and metallicities, and we assume [Mg/Fe] to be a proxy for the [$\alpha$-elements/Fe] ratio. The relative nature of this method to derive the abundance ratio might not perfectly capture the absolute [Mg/Fe] value, but it is well suited for a relative comparison as intended in this Letter, since it minimizes the sensitivity of our results to the unavoidable degeneracies. 

\section{Results}

To investigate the influence of the BH on the stellar population properties of the host galaxies, we divide our sample into over-massive BH galaxies (i.e., galaxies lying above the BH mass -- $\sigma$ relation) and under-massive BH galaxies (Fig.~\ref{sigma}), using the best-fitting relation derived in \citet{vdB}.

In the local Universe, the stellar population properties of galaxies correlate with their central $\sigma$ \citep[e.g.][]{Peletier,Thomas,Labarbera}, which is generally interpreted as a galaxy-mass driven evolution. In order to remove these trends from the analysis, we study the variations in stellar population properties at fixed $\sigma$, dividing our sample of galaxies into four bins. For each of these bins we stack the spectra of over- and under-massive BH galaxies, and we calculate their (averaged) stellar population properties. 

\begin{figure}
 \begin{center}
  \includegraphics[scale=0.45]{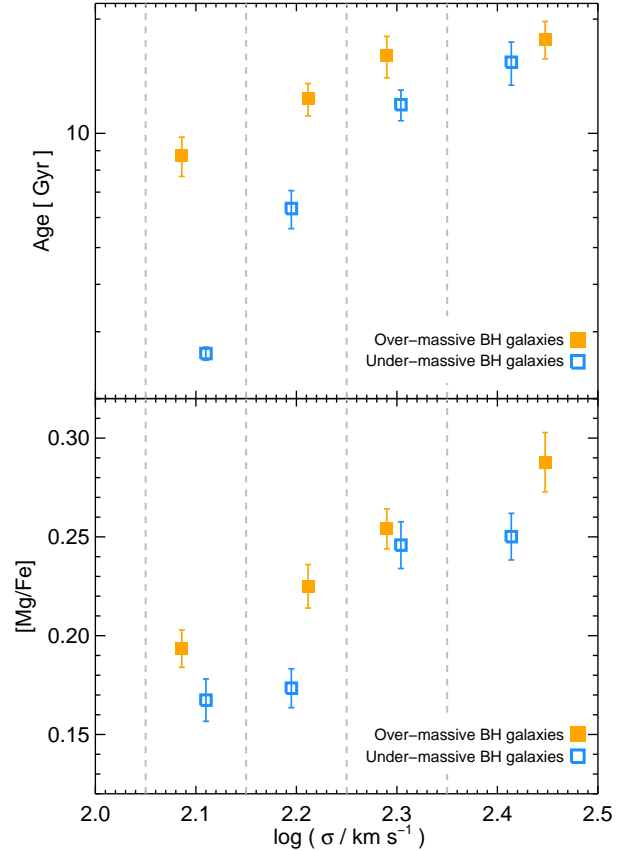}
  \caption{Age (top) and [Mg/Fe] (bottom) variations as a function of galaxy velocity dispersion for over- (orange) and under-massive (blue) BH galaxies. Dashed vertical lines indicate the different $\sigma$ bins, and measurements come from our stacked spectra. Two main trends are shown in this figure. First, as expected, the age and the [Mg/Fe] ratio increase with $\sigma$. Second, the stellar population properties depend on the central BH mass, being older and more $\alpha$-enhanced for galaxies which, at fixed $\sigma$, host a more massive BH.}
  \label{pop}
 \end{center}
\end{figure}

Fig.~\ref{pop} shows the best-fitting age and [Mg/Fe] values in our four $\sigma$ bins, for over- and under-massive BH galaxies. In general, galaxies become older and more $\alpha$-enhanced with increasing $\sigma$. The fact that we recover the expected trends with galaxy mass reinforces the consistency of our stellar population analysis. In addition to this well-known dependence with $\sigma$, we observe a clear difference depending on the mass of the BH: {\it galaxies which at fixed $\sigma$ have more massive BHs, also tend to be older and more $\alpha$-enhanced}.

The decreasing age difference with increasing galaxy mass can be understood as a combination of two factors. First of all, line-strength analysis provides luminosity weighted values, with ages being more sensitive to this bias than are metallicity and $\alpha$-enhancement \citep{Serra}. Thus, although the relative differences in age might be similar along the whole sample, they appear more prominent when probing lower mass and therefore younger galaxies. Note that a small fraction of young stars could boost these luminosity weighted measurements, which therefore should be considered as upper limits for the age difference. Second, as the stellar populations get older, our precision in the age determination decreases. Consequently, as we consider the massive end of our sample, the ages of the galaxies tend to be indistinguishable.

Differences in the [Mg/Fe] ratio are less dependent on $\sigma$. Note that for low mass galaxies, the [Mg/Fe] ratio is also biased towards young stellar populations, which by definition result from more extended star formation histories and therefore are less $\alpha$-enhanced. The $\sigma\sim2.3\log \mathrm{km\,s}^{-1}$ bin shows a similar [Mg/Fe] value for under- and over-massive BH galaxies. We suggest this is due to a small range in the BH masses for that particular bin ($\sim0.5$ dex, compared to a typical $\sim1$ dex for the other bins), which is apparent in Fig.~\ref{sigma}.

\begin{table}
\centering
\begin{center}
\begin{tabular}{l c c || l c c}
\hline
Galaxy & Age & [Mg/Fe] & Galaxy & Age & [Mg/Fe]\\
       & (Gyr) &  (dex) &           & (Gyr)  & (dex) \\
\hline
NGC\,0307   &  13.2   &    0.18     &    NGC\,3842   &  16.0   &    0.31  \\
NGC\,0315   &  19.7   &    0.17     &    NGC\,3953   &  7.0    &    0.10  \\
NGC\,0383   &  17.5   &    0.20     &    NGC\,4026   &  14.9   &    0.13  \\
NGC\,0524   &  12.2   &    0.21     &    NGC\,4143   &  17.0   &    0.20  \\
NGC\,0541   &  16.7   &    0.19     &    NGC\,4203   &  19.7   &    0.14  \\
NGC\,0741   &  17.8   &    0.21     &    NGC\,4261   &  12.6   &    0.21  \\
NGC\,0821   &  10.1   &    0.16     &    NGC\,4335   &  8.9    &    0.17  \\
NGC\,1023   &  16.8   &    0.18     &    NGC\,4350   &  9.7    &    0.18  \\
NGC\,1271   &  19.5   &    0.23     &    NGC\,4459   &  3.0    &    0.14  \\
NGC\,1277   &  18.6   &    0.32     &    NGC\,4473   &  11.6   &    0.18  \\
NGC\,1961   &  5.5    &    0.15     &    NGC\,4564   &  8.9    &    0.20  \\
NGC\,2892   &  12.6   &    0.19     &    NGC\,4649   &  15.7   &    0.33  \\
NGC\,2960   &  1.4    &    0.00     &    NGC\,4698   &  16.4   &    0.18  \\
NGC\,3115   &  8.3    &    0.22     &    NGC\,5127   &  11.3   &    0.20  \\
NGC\,3245   &  11.2   &    0.15     &    NGC\,5490   &  16.0   &    0.24  \\
NGC\,3368   &  2.07   &    0.06     &    NGC\,5576   &  3.6    &    0.12  \\
NGC\,3377   &  9.8    &    0.24     &    NGC\,6086   &  19.4   &    0.25  \\
NGC\,3379   &  17.5   &    0.24     &    NGC\,7052   &  14.4   &    0.20  \\
NGC\,3414   &  15.7   &    0.19     &    NGC\,7331   &  2.6    &    0.20  \\
NGC\,3607   &  8.0    &    0.20     &    NGC\,7619   &  7.4    &    0.29  \\
NGC\,3608   &  14.7   &    0.22     &    NGC\,7768   &  18.9   &    0.27  \\
NGC\,3627   &  1.3    &    0.10     &    MRK\,1216   &  19.6   &    0.24  \\
\hline               
\end{tabular}
\caption{Best-fitting ages and [Mg/Fe] ratios. We include here only galaxies where the signal-to-noise allowed a reliable stellar population analysis. Ages older than 14 Gyr are due to very low H$\beta$ values and the fact that we assumed a universal IMF slope \citep{Labarbera}.}
\label{tab:fit}
\end{center}
\end{table}

\begin{figure}
 \begin{center}
  \includegraphics[scale=0.45]{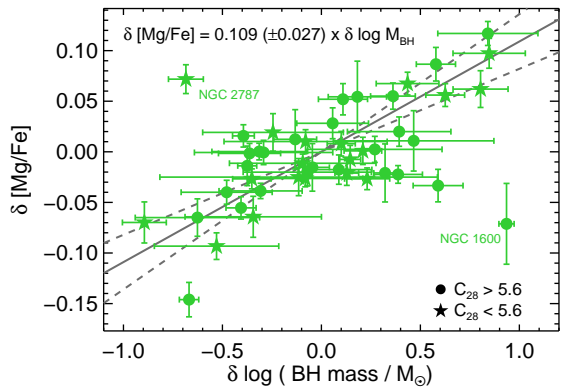}
  \caption{[Mg/Fe] vs.\ BH mass for individual galaxies. The residuals from the [Mg/Fe]--$\sigma$ relation correlate positively with the scatter in the BH mass--$\sigma$ relation. The solid line corresponds to the best-fitting relation, and dashed lines indicate the uncertainty in the fit. At a given $\sigma$, the $\alpha$-enhancement of a galaxy seems to follow the mass of its central BH. Filled circles and stars indicate high and low concentration indices, respectively.}
  \label{scatter}
 \end{center}
\end{figure}

As an additional test we compare individual BH masses and stellar population measurements, listed in Table~\ref{tab:fit}. As stated above, both BH masses and stellar population properties correlate with $\sigma$, so this galaxy mass-dependency has to be corrected to isolate the potential BH effect. In Fig.~\ref{scatter}, we show the residuals in the BH mass--$\sigma$ relation plotted against the residuals of the [Mg/Fe]--$\sigma$ relation, for individual galaxies. After removing the main correlation with $\sigma$, those galaxies hosting more massive BHs also exhibit more  $\alpha$-enhanced stellar populations. Note that Fig.~\ref{scatter} shows measurements over a wide range of log~$\sigma$ (2.0--2.5), further supporting the averaged trend shown in Fig.~\ref{pop}. Two clear outliers depart from the global trend. On the upper left corner of Fig.~\ref{pop}, NGC~2787 has a pseudo-bulge with an enhanced [Mg/Fe] ratio ($+$0.26 dex) and a relatively low-mass BH (10$^{7.6}$\msun), pointing towards a non-AGN related quenching process. More interesting is the other outlier, NGC~1600, a galaxy with a very massive BH but showing a relatively mild $\alpha$ enhancement. Either the [Mg/Fe] or the BH mass measurements of NGC~1600 are incorrect or the present-day BH mass was reached after the bulk of the star formation took place in the host galaxy. 

\begin{figure*}
 \begin{center}
  \includegraphics[scale=0.5]{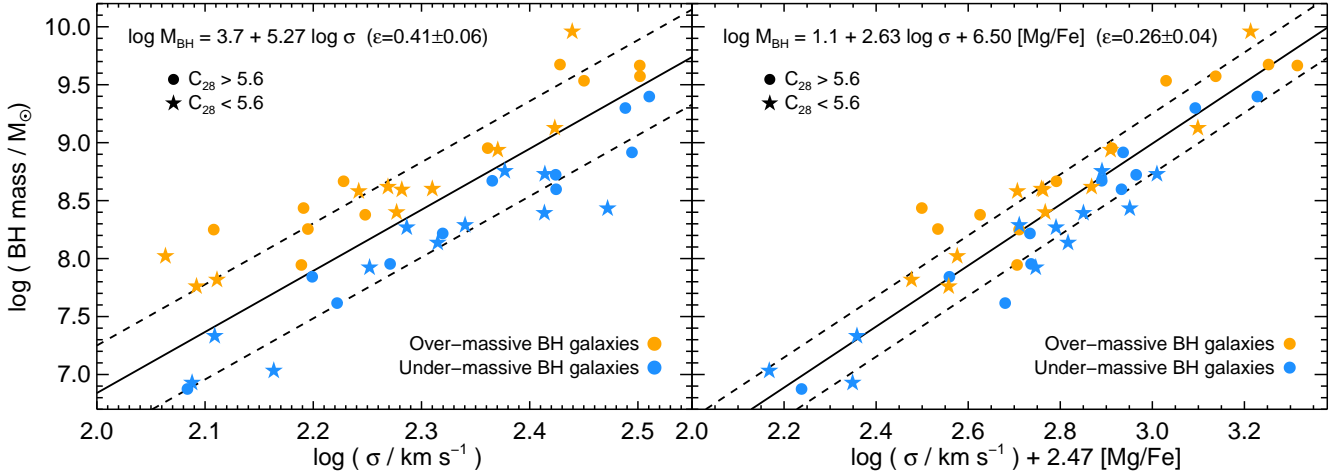}
  \caption{Best-fitting relations between observed BH masses, velocity dispersion, and [Mg/Fe]. The solid line corresponds to the best-fitting solution, and dashed lines indicate the intrinsic scatter. Orange and blue symbols indicate over- and under-massive BH galaxies, and filled circles and stars indicate high and low concentration indices, respectively. The left panel shows the reference BH mass--$\sigma$ relation, whereas in the right panel we fitted a regression of the form $\log \mathrm{M_{BH}} = \mathrm{constant} + \alpha\log\sigma+\beta\mathrm{[Mg/Fe]}$, which has a significantly lower scatter. Neither of the two panels includes the two outliers of Fig.~\ref{scatter}.}
  \label{relation}
 \end{center}
\end{figure*}

Finally, in Fig.~\ref{relation} we make use of the correlation, at fixed $\sigma$, between BH mass and [Mg/Fe], to revisit the fundamental BH mass--$\sigma$ relation. Although $\sigma$ is known to be the main and almost only relevant parameter for determining the BH mass \citep{Beifiori,vdB}, we found that also taking into account the [Mg/Fe] values significantly reduces the scatter, from $\epsilon=0.41\pm0.06$ to $\epsilon=0.26\pm0.04$. To be consistent with the analysis presented in \citet{vdB}, we used the Bayesian routines of \citet{Kelly} to estimate both the scatter and the best-fitting relations shown in  Fig.~\ref{relation}.

\section{Discussion and conclusions}

Understanding the mechanism responsible for quenching the star formation within massive galaxies is a difficult task. In the local Universe, a detailed analysis of their stellar populations is possible, but quenching is no longer taking place. At higher redshifts ($z\gtrsim2-3$), when massive galaxies ceased forming new generations of stars, obtaining sufficiently good spectra is out of reach for the current generation of telescopes.

In this Letter we made use of two observables left behind in the natural evolution of galaxies: the BH as power source of the AGN activity, and the abundance pattern of stars as a proxy for the formation time-scale of the stellar populations. Our findings, summarized in Figs.~\ref{pop} and \ref{scatter}, indicate a strong connection between the stellar population properties of a galaxy and the mass of its central BH. In particular, stars within over-massive BH galaxies tend to be older and more $\alpha$-enhanced, suggesting a more rapid quenching process than in under-massive BH galaxies.

It could be argued that our findings result from spurious correlations with unrecognized parameters. However, \citet{Beifiori} and \citet{vdB} have shown that the residuals of the BH mass--$\sigma$ relation do not correlate with structural parameters of galaxies.

An alternative scenario could involve a sample biased towards later-type galaxies below the BH mass--$\sigma$ relation \citep{Kormendy}. Nevertheless, we rule out this possibility, first of all, because the observed trend also extends to the most massive galaxies in our sample, where only early-type galaxies are found. In addition, after removing the pseudo-bulges from the analysis, the same trends with BH mass are recovered. \citet{Greene} recently claimed that megamaser disk galaxies are slightly off-set from the main BH mass--$\sigma$ relation. Most of these objects were removed because their spectra displayed strong nebular emission within the H$_\beta$ line, and only one megamaser (NGC~2960) is present in the final sample. As expected from our findings, and given its relatively low BH mass, NGC~2960 shows almost no $\alpha$-enhancement ([Mg/Fe] = 0.003)

Notice that, although not ideal, our approach for correcting the stellar continuum does not introduce any bias to our stellar population analysis, since it is completely independent of the mass of the BH. In principle, the mass of the BH in $\alpha$-enhanced galaxies could be over-estimated if a solar-scaled ([Mg/Fe] = 0) mass-to-light ratio (M/L) is assumed. However, the effect of the abundance pattern on the M/L is very mild, and only significant for filters bluer than $\lambda_\mathrm{eff}\sim4000$ \AA \ \citep{alpha}. Moreover, since dynamical models commonly assume a radially constant M/L, a pronounced gradient may also affect the BH measurement. In this regard, \citet{Pat} have shown that more massive and  $\alpha$-enhanced ETGs galaxies have flatter gradients in their stellar population properties, and thus, more constant M/L profiles \citep{mn15}. Thus, we do not expect a systematic bias in the BH masses due to the stellar population properties of the host galaxies.

We interpret our results as a direct connection between the central BH and the star formation history of galaxies. More massive BHs would have formed earlier and in denser regions of the Universe, feeding more active AGNs, and thus quenching the star formation more quickly. This scenario leads to older and more $\alpha$-enhanced stellar populations. As the BH mass and the AGN feedback decrease, galaxies would form later, creating stars over more extended periods of time, leading to lower [Mg/Fe] and younger ages. For galaxies above L$^*$, like those studied in this work, the effect of stellar feedback even in the early stages of galaxy formation is expected to be negligible. However, a natural prediction of this AGN-driven quenching is that the differences between over- and under-massive galaxies would start vanishing for objects below  L$^*$.

Is then the chemical enrichment of massive galaxies entirely determined by the AGN activity in the early Universe? The abundance pattern of a galaxy depends on two factors: the formation time-scale of its stellar populations and the number of massive stars responsible for the chemical enrichment, i.e., the stellar initial mass function (IMF). Consequently, an enhanced [Mg/Fe] can result from a short star-formation event or from a more extended star-formation history but with a flat (giant-dominated) IMF \citep{vazdekis96,thomas99}. While our analysis at fixed $\sigma$ shows an enhanced [Mg/Fe] for over-massive BH galaxies, therefore supporting a connection between AGN and the abundance pattern, the IMF might also be playing an important role in establishing the [Mg/Fe]--$\sigma$ relation. In particular, it has been shown that a time-varying IMF, which is flatter at earlier epochs, is necessary to reconcile the chemical properties of nearby massive galaxies and their apparently non-universal IMF \citep{Ferreras15,weidner:13,mn16}. Numerical simulations have also supported the idea of a flat IMF during the early formation of massive galaxies \citep{Calura,Arrigoni,Fontanot}. Thus, the [Mg/Fe]--$\sigma$ relation is potentially driven by a combination of the two processes,  AGN-related quenching plus a non-universal IMF.

Irrespective of the origin of the abundance pattern in galaxies, {\it our results demonstrate, observationally, a strong correlation between the black hole masses and the star formation histories of galaxies, which we interpret as AGN feedback directly driving the star formation history of massive galaxies}.

\acknowledgments 
We acknowledge support from NSF grant AST-1211995. IMN would like to thank Jes\'us Falc\'on-Barroso, Luis Peralta de Arriba and Marja Seidel for their useful comments and suggestions.

\bibliographystyle{apj}
% \bibliography{hetmgs}

\end{document}